\documentclass[10pt,conference]{IEEEtran}

\AtBeginDocument{
  
}

\usepackage{listings}
\usepackage{tcolorbox}
\usepackage{soul}
\usepackage[dvipsnames]{xcolor}
\usepackage{fancybox}
\usepackage{xspace}
\usepackage{ragged2e}
\usepackage{changepage}
\usepackage{makecell}
\usepackage{setspace}
\usepackage{enumitem}
\usepackage{booktabs, multirow}
\usepackage{pgfplots}
\pgfplotsset{compat=1.18}
\usepackage{cite}

\usepackage{url}

\usepackage{hyperref}
\hypersetup{
  colorlinks,
  citecolor=blue,
  linkcolor=blue,
  urlcolor=blue,
}

\definecolor{codebg}{RGB}{248,248,248}
\definecolor{codekw}{RGB}{0,92,197}
\definecolor{codestr}{RGB}{163,21,21}
\definecolor{codecmt}{RGB}{0,128,0}
\definecolor{codeln}{RGB}{120,120,120}

\lstdefinestyle{logcode}{
  backgroundcolor=\color{codebg},
  basicstyle=\ttfamily\tiny,
  keywordstyle=\color{codekw}\bfseries,
  stringstyle=\color{codestr},
  commentstyle=\color{codecmt}\itshape,
  numbers=left,
  numberstyle=\tiny\color{codeln},
  numbersep=6pt,
  stepnumber=1,
  showstringspaces=false,
  breaklines=true,
  frame=single,
  rulecolor=\color{black!20},
  tabsize=2,
  captionpos=b
}

\usepackage{amsmath,amssymb,amsfonts}

\usepackage{textcomp}
\usepackage{lipsum}
\usepackage{graphicx}
\usepackage{tikz}
\usetikzlibrary{positioning,arrows.meta,shapes.misc,fit,calc}
\usepackage{subfigure}
\usepackage{balance}
\usepackage{wrapfig}
\usepackage{needspace}
\usepackage[ruled, lined, linesnumbered, commentsnumbered, longend]{algorithm2e}
\usepackage{algorithmic}
\SetCommentSty{mycommfont}
\usepackage{array}

\usepackage{adjustbox}
\usepackage{threeparttable}

\newcommand{\phead}[1]{\vspace{1mm} \noindent {\bf #1}}
\newcommand{\methodname}{LogSemFuse\xspace}
\newcolumntype{?}{!{\vrule width 1.5pt}}

\newcommand{\rqboxc}[1]{
  \begin{tcolorbox}[left=1pt,right=1pt,top=1pt,bottom=1pt,colback=gray!5,colframe=gray!40!black,before skip=5pt,after skip=0pt]#1
\end{tcolorbox}}

\definecolor{lightgray}{gray}{0.9}

\makeatletter
\newcommand*{\rom}[1]{\expandafter\@slowromancap\romannumeral #1@}
\makeatother

\begin{document}

\title{\methodname: Semantic Evidence Fusion for Explainable Log Anomaly Detection}
\author{
\IEEEauthorblockN{Hassan Jabri\textsuperscript{\dag}, Zeyang Ma\textsuperscript{\dag *}, Zhijie Wang\textsuperscript{\S}, Tse-Hsun (Peter) Chen\textsuperscript{\dag}}
\IEEEauthorblockA{\textsuperscript{\dag}Software PErformance, Analysis and Reliability (SPEAR) Lab, Concordia University, Montreal, Quebec, Canada\\
\textsuperscript{\S}Concordia University, Montreal, Quebec, Canada\\
hajabri95@gmail.com \quad m\_zeyang@encs.concordia.ca \quad zhijie.wang@concordia.ca \quad peterc@encs.concordia.ca}
}
\maketitle
\begingroup
\renewcommand{\thefootnote}{*}
\footnotetext{Zeyang Ma is the corresponding author.}
\endgroup

\begin{abstract}
Log anomaly detection is critical for reliability monitoring and failure diagnosis in modern software systems.
Existing model-based detectors provide useful anomaly signals, but they can still miss anomalous sessions and typically expose only scores or labels rather than the operational semantics behind a decision.
This lack of semantic evidence limits their ability to explain why a session is anomalous, even when the final anomaly label is correct.
The gap matters in practice because operators need to distinguish urgent failures from benign deviations and trace suspicious sessions back to concrete operational behavior.
LLMs can recover richer log semantics, but using them as standalone detectors or repeatedly generating free-form explanations can be costly and difficult to reuse.
We present \methodname, an evidence-guided plug-in framework that enhances existing backbone detectors without replacing their original pipelines.
\methodname combines backbone predictions with reusable semantic evidence from local event patterns, LLM-based semantic reasoning, and cluster-derived executable rules to produce both anomaly decisions and evidence-based explanations.
The resulting output reports the final label together with the semantic evidence that supports it, such as fired local patterns, triggered rules, and LLM rationale.
We evaluate \methodname on HDFS, BGL, and Liberty using DeepLog, LogAnomaly, LogBERT, and NeuralLog as backbones.
\methodname improves every non-perfect baseline, preserves the already perfect case, recovers 98.8\% of backbone false negatives, and produces explanations preferred over direct LLM explanations in a human study.
These gains require only modest and stable inference-time overhead, showing that semantic augmentation can improve detection effectiveness and interpretability without imposing large runtime costs.
\end{abstract}


\section{Introduction}

Modern software systems generate large volumes of logs that record execution states, component interactions, warnings, and failures.
These logs are widely used for system monitoring, failure diagnosis, and reliability engineering, making log anomaly detection a critical software engineering task~\cite{he2016experience,he2021survey,zhu2023loghub}.
However, logs are often lengthy, repetitive, and difficult to interpret manually.
To automate log analysis, log anomaly detection techniques commonly parse raw messages into templates, map templates to EventIds, group events into sessions, and classify each session as normal or anomalous~\cite{du2016spell,he2017drain,he2021survey,khan2024impact}.

Existing model-based detectors learn statistical, sequential, contextual, or semantic patterns from these logs and achieve strong anomaly detection performance~\cite{du2017deeplog,meng2019loganomaly,guo2021logbert,zhang2019robust,yang2021plelog,le2021neuralog}.
These detectors typically output only an anomaly score or label, without explaining which log evidence supports the decision or what abnormal operational behavior it represents.
Operators must therefore inspect the logs themselves to understand why a session was flagged, limiting the usefulness of detection results for diagnosis and incident response~\cite{liao2020questioning,zhang2020effect,wang2023deepseer}.

General-purpose XAI methods, such as LIME~\cite{ribeiro2016lime} and SHAP~\cite{lundberg2017shap}, mainly provide feature-level attributions and do not directly explain log-specific operational behavior.
Generating such explanations requires connecting low-level log evidence, including raw messages, templates, and EventId sequences, to the higher-level operational behavior represented by the session.
Large language models (LLMs) provide a promising way to interpret such semi-structured evidence, and recent work has explored their use for log anomaly detection and log interpretation~\cite{qi2023loggpt,zhang2023logprompt,liu2024logprompt,guan2024logllm}.
However, directly prompting an LLM does not ensure that its explanation is grounded in the evidence underlying the anomaly decision~\cite{madsen2024selfexplanations}.

At the same time, applying an LLM directly to every log session can be costly and may not be practical for large-scale deployment~\cite{openai2024gpt4omini}.
Meanwhile, existing model-based log anomaly detectors already achieve strong detection performance on many benchmarks.
This motivates a different design: rather than replacing existing detectors, can LLMs be used selectively to provide complementary reasoning and explanatory capabilities while keeping inference costs low?

In this paper, we propose \methodname.
\methodname is a plug-in framework: it does not replace the backbone detector or change its training objective, but augments the backbone decision with structured semantic evidence derived from log events, templates, LLM reasoning, and anomaly clusters identified during training. 
Specifically, \methodname uses three semantic evidence modules: (1) a local event-pattern matcher that mines discriminative unigrams, bigrams, and trigrams from labeled training sessions; (2) an LLM-based semantic reasoning module that produces a high-risk prediction, confidence score, and expected-versus-observed rationale for each session; and (3) a cluster-derived rule construction module that converts LLM analyses of anomalous training-session clusters into rules.

The event patterns and cluster-derived rules learned during training are stored in an event-pattern bank and a rule bank, respectively, to support their efficient retrieval and reuse across sessions. Both knowledge banks remain fixed at inference time, while session-level LLM reasoning results are cached by session identifier to avoid redundant inference. For each incoming session, \methodname combines four binary signals from the backbone, local event-pattern matcher, cluster-derived rule bank matcher, and LLM-based semantic reasoning module using a fusion operator selected on the validation set. It then reports the matched patterns, triggered rules, and LLM rationale as evidence supporting the final decision.

We evaluate \methodname on HDFS, BGL, and Liberty using DeepLog, LogAnomaly, LogBERT, and NeuralLog as backbone detectors~\cite{du2017deeplog,meng2019loganomaly,guo2021logbert,le2021neuralog}.
\methodname improves every non-perfect baseline while preserving the already perfect case, and recovers 98.8\% of backbones' false negatives across test sets.
Furthermore, a within-subject user study ($N=14$) shows that \methodname can generate high-quality explanations compared to a baseline approach.
Participants rated the explanations from \methodname higher than the baseline LLM in overall quality, with mean scores of 4.13 versus 3.06 on a five-point Likert scale.
With modest and stable inference-time overhead, \methodname improves detection effectiveness, provides evidence-based explanations, and amortizes LLM usage through reusable knowledge-bank artifacts, making it a practical framework for semantic augmentation.

The paper makes the following contributions:
\begin{itemize}
    \item We propose \methodname, a plug-in semantic augmentation framework that enhances existing backbone log detectors with executable semantic evidence while preserving their original training and inference pipelines.

    \item We design a knowledge bank that stores reusable local event patterns and cluster-derived rules, while caching session-level LLM-based semantic reasoning results to enable efficient and low-cost reuse.
    \item We design an evidence-based explanation approach for log anomaly detection that explains each anomaly decision using fired local event patterns, triggered cluster-derived rules, and LLM-based semantic reasoning results.
    \item We evaluate \methodname across

    three log anomaly detection datasets and four backbone detectors, showing that it improves or preserves detection performance with modest and stable runtime overhead and LLM cost, while providing high quality explanations for human developers.

\end{itemize}

\noindent\textbf{Paper Organization.}
Section~\ref{sec:background-related} formulates the task and reviews related work.
Section~\ref{sec:methodology} presents the \methodname framework.
Section~\ref{sec:setup} describes the experimental setup, and Section~\ref{sec:evaluation} reports the evaluation results.
Section~\ref{sec:threats-to-validity} discusses threats to validity.
Section~\ref{sec:conclusion} concludes the paper.

\section{Background and Related Work}
\label{sec:background-related}

This section formulates the log anomaly detection problem and positions \methodname against related work.

\subsection{Background}
\label{subsec:background}

\phead{Log-based Anomaly Detection.}
System logs record execution states, resource events, warnings, errors, and component interactions, making them a common data source for reliability monitoring and failure diagnosis.
Because raw logs contain variable fields such as identifiers, paths, timestamps, and sizes, log anomaly detection pipelines usually first parse messages into templates and assign each template an EventId~\cite{he2016experience,he2021survey,khan2024impact}.
The parsed events are first ordered chronologically and grouped into sessions according to the dataset semantics. Events sharing an execution identifier, such as a block, request, or trace ID, are grouped into one session.

For example, \([E_3,E_7,E_7,E_{12},E_5]\) may form a session because the events have the same block ID. When no such identifier is available, sessions are constructed using rules such as fixed-length windows.
Most log anomaly detection models operate on the parsed EventId or template sequence associated with each session\cite{du2017deeplog,meng2019loganomaly,guo2021logbert,le2021neuralog}. From the training logs, they learn recurring sequential, co-occurrence, or contextual patterns. At inference time, they identify sessions whose event patterns deviate from those observed during training.

\phead{Current Challenge.} Despite strong detection performance, many model-based anomaly detectors provide only an anomaly score or label, with limited evidence explaining the decision. Although log templates often retain operationally meaningful information, detectors typically model these templates as symbolic inputs without translating their predictions back into such semantics. Operators may therefore know that a session is anomalous but not which behavior caused the alert or how it supports triage and debugging. This limitation motivates techniques that complement existing detectors with interpretable semantic evidence.

\subsection{Related Work}
\label{subsec:related-work}

\phead{Model-based Log Anomaly Detection.}
Many log anomaly detectors rely on parsed representations produced by log parsers such as Spell~\cite{du2016spell} and Drain~\cite{he2017drain}, and recent work studies how parsing quality affects downstream anomaly detection~\cite{zhu2023loghub,jiang2024logparsing,khan2024impact}.
Classical methods model statistical regularities, event counts, or invariants~\cite{xu2009detecting,lou2010mining}, while clustering methods such as LogCluster~\cite{lin2016logcluster} group related log sequences for problem identification.
Deep learning-based methods learn richer temporal or contextual patterns from event sequences, including recurrent, feature-enhanced, Transformer-based, graph-based, parser-free, and pretraining-based approaches~\cite{du2017deeplog,meng2019loganomaly,guo2021logbert,huang2020hitanomaly,xie2022loggd,le2021neuralog,guo2024logformer}.
Hybrid detectors such as LogRobust and PLELog further incorporate semantic representations to improve robustness under unstable or weakly labeled logs~\cite{zhang2019robust,yang2021plelog}.
\methodname complements these detectors because it treats them as backbones rather than replacing their architecture, input representation, or training pipeline.
\methodname's semantic artifacts are fused at the decision level and returned as evidence supporting the explanation, including matched local event patterns, triggered cluster-derived rules, LLM-based semantic reasoning results, and cluster-level explanations.

\phead{LLM-based Log Anomaly Detection and Diagnosis.}
Recent work explores LLMs for log anomaly detection and online log interpretation.
LogGPT~\cite{qi2023loggpt} investigates ChatGPT-style models for log anomaly detection, LogPrompt studies prompt strategies for anomaly detection and interpretable online log analysis~\cite{liu2024logprompt}, and LogLLM~\cite{guan2024logllm} trains an LLM-based framework for log anomaly detection.
These approaches show that LLMs can surface useful log semantics that may not be exposed by EventId-only detector outputs.
At the same time, reported detection results for LLM-based detectors remain uneven.
For example, LogGPT reports an F1-score of 0.618 on BGL under its default content-sequence prompt, while LogPrompt reports an F1-score of 0.384 on BGL under its CoT prompt strategy~\cite{qi2023loggpt,liu2024logprompt}.
In contrast, traditional detectors such as DeepLog~\cite{du2017deeplog} and LogAnomaly~\cite{meng2019loganomaly} model stable sequential and quantitative log patterns and have been reported to achieve BGL F1-scores above 0.9 in prior benchmarks.
This contrast suggests that LLM-centric detection can benefit from being combined with traditional detector signals rather than replacing them.
However, LLM-centric detection raises concerns about inference cost and response variability.
\methodname instead uses LLM-based semantic reasoning as one semantic evidence module, while the backbone remains responsible for sequence-based anomaly detection.

\phead{Explaining Log Anomalies and Rule-based Evidence.} Previous studies have shown that explainable AI (XAI) can improve user trust and confidence in AI~\cite{liao2020questioning, zhang2020effect, wang2023deepseer}.
General-purpose XAI methods such as LIME~\cite{ribeiro2016lime} and SHAP~\cite{lundberg2017shap} explain model predictions through local surrogate models or feature attributions. Systems like DeepSeer~\cite{wang2023deepseer} show the value of interactive explanations for sequence models.
However, these methods are not designed around log-specific artifacts such as EventId sequences, templates, session-level failure semantics, or reusable operational rules.

Rule-based log analysis, such as LogRules~\cite{huang2025logrules}, is relevant to \methodname because a matched rule identifies both the event pattern that triggered the alert and its operational meaning.
However, \methodname does not explain a prediction from rule matches alone. It also compares the target session with similar historical log sequences, integrates evidence from its detection signals, and uses an LLM to synthesize this evidence into a session-specific semantic explanation. The findings become more natural and easier for developers to understand.

\noindent
\textbf{Position of \methodname.}
Compared with prior model-based, LLM-based, rule-based, and explainability-oriented approaches, \methodname is best understood as a semantic augmentation framework for existing log anomaly detectors.
Its contribution is not a new standalone detector, LLM-centered rule repository, or post-hoc explanation tool, but a unified way to enhance backbone detectors with reusable semantic evidence while producing evidence-based semantic explanations.

\section{Methodology}
\label{sec:methodology}

\begin{figure*}[t]
  \centering
  \includegraphics[width=1.8\columnwidth]{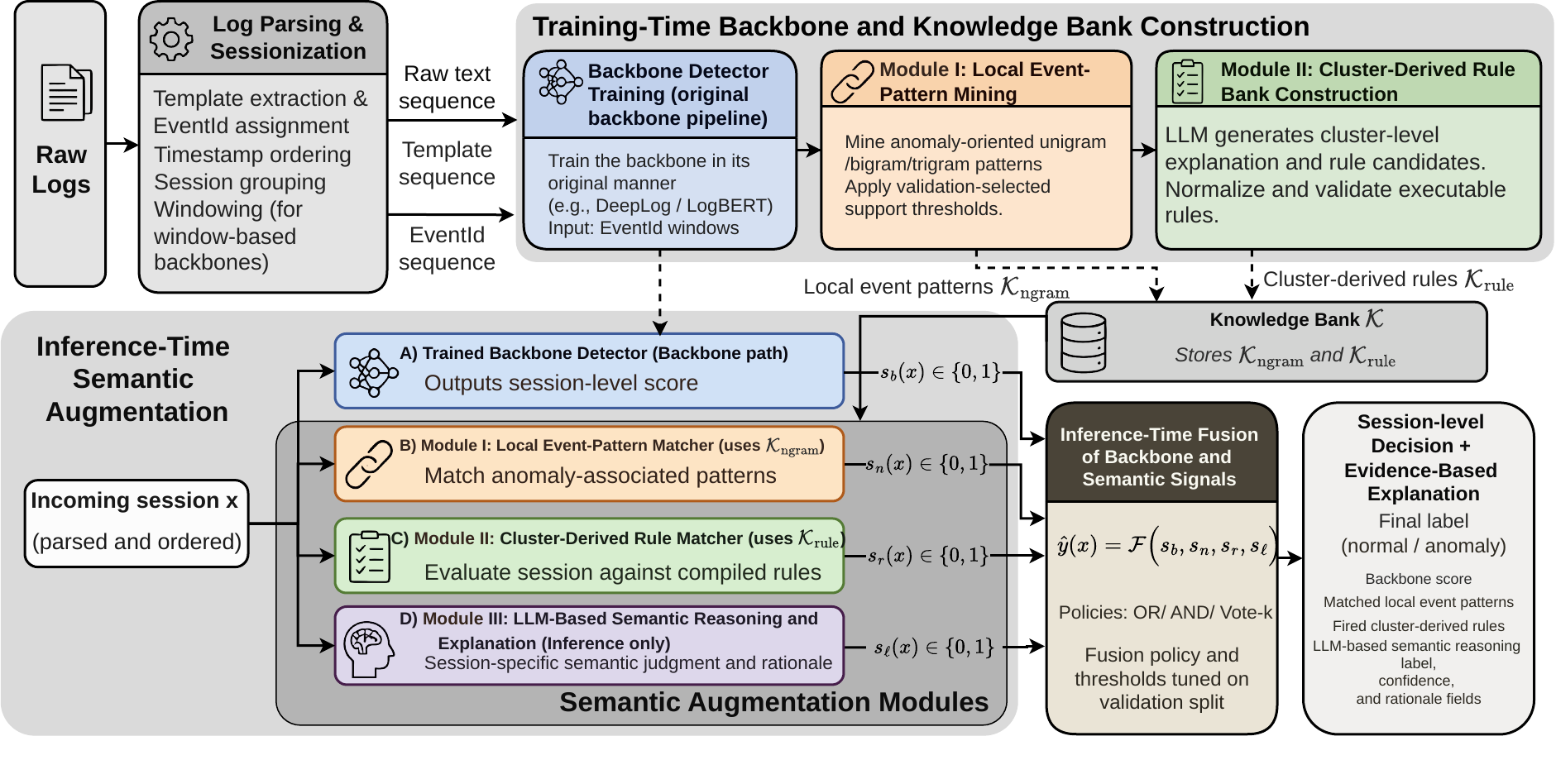}

  \vspace{-2em}
  \caption{Overview of \methodname pipeline.}
  \label{fig:pipeline_overview}
  \vspace{-1.5em}
\end{figure*}

\methodname is a plug-in framework designed to improve the accuracy and explainability of existing log anomaly detectors by incorporating semantic evidence from log sessions. Rather than modifying or replacing the detector, \methodname preserves the backbone's original training and inference pipeline. It combines backbone prediction with semantic evidence to improve final anomaly detection and to explain why a session is classified as anomalous.

Figure~\ref{fig:pipeline_overview} illustrates the workflow.
Raw logs are parsed into templates, ordered by timestamp, and grouped into sessions.
The backbone detector is trained and executed using its original pipeline.
Separately, \methodname constructs \emph{reusable evidence} from the training data and stores it in a knowledge bank \(\mathcal{K}\).
At inference time, an incoming session is evaluated by the backbone, matched against reusable evidence, and analyzed through LLM-based semantic reasoning. These signals are fused to produce a final session-level prediction together with evidence supporting the decision and its explanation.

\subsection{Input Representation and Backbone Interface}
\label{sec:representation}

\phead{Log Representation.}
\methodname converts raw logs into structured event sequences using a standard preprocessing pipeline~\cite{he2016experience,he2021survey}.
First, log lines are parsed into templates using Drain~\cite{he2017drain}.
Each template is assigned a unique \textit{EventId}, which transforms raw log text into a discrete event sequence.
Variable fields, such as identifiers, sizes, and file paths, are replaced with placeholders to support consistent template matching across sessions.
We then order log entries by timestamp and group them into sessions using dataset-specific identifiers, such as block IDs or request IDs~\cite{du2017deeplog,he2021survey,ma2026llm4log}.
Each session represents a coherent execution trace.
The detailed sessionization rules for each dataset are described in Section~\ref{sec:datasets}.

For window-based backbone detectors, \methodname constructs fixed-length windows over the event sequence following prior sequence-based log anomaly detectors~\cite{du2017deeplog,meng2019loganomaly}.
Window-level predictions are later aggregated into session-level decisions to ensure a consistent evaluation granularity across all detectors.
For each session, \methodname maintains three aligned views of the same log behavior: raw log text for human-readable tracing, template sequences for semantic analysis, and EventId sequences for backbone modeling and deterministic evidence matching.

\phead{Backbone Interface.}
\methodname performs semantic augmentation during inference-time decision making.
The backbone is trained and executed on EventId sequences using its original architecture, input representation, and learning objective.
After the backbone produces an anomaly score or label, \methodname fuses its output with signals from the local event-pattern matcher, cluster-derived rule bank matcher, and LLM-based semantic reasoning module.
Keeping these
semantic evidence modules outside backbone training preserves compatibility with existing detectors while adding evidence for final decision and its explanation.

\subsection{Training-Time Reusable Evidence Construction}
\label{sec:evidence_construction}
\methodname constructs reusable evidence from training logs before inference. This evidence captures recurring local patterns and higher-level operational behaviors that help identify and explain anomalous sessions. At inference, \methodname matches this evidence to a new session to support its anomaly prediction. Let \(x\) denote a log session, and let \(E(x)\) denote its sequence of event identifiers. \methodname builds a \textit{\textbf{knowledge bank}} \(\mathcal{K}\) containing two types of evidence: a \textit{\textbf{local event-pattern bank}} \(\mathcal{K}_{\text{ngram}}\) and a \textit{\textbf{cluster-derived rule bank}} \(\mathcal{K}_{\text{rule}}\). The local event-pattern bank captures short, discriminative event patterns, whereas the cluster-derived rule bank captures broader recurring operational behaviors summarized from groups of similar anomalous sessions. Both types of evidence are constructed exclusively from the training split and are matched deterministically against new sessions at inference.

\subsubsection{\textbf{Local Event-Pattern Mining}}
\label{sec:ngram_detector}

Local event patterns provide simple but interpretable evidence for abnormal execution.
Many log anomalies appear through short event combinations, such as repeated retries, exception-following operations, or unusual local transitions around a failure point~\cite{landauer2023critical}.
Although neural detectors may learn such patterns implicitly, explicit event-pattern evidence is useful because it is deterministic, reusable, and directly inspectable~\cite{mantyla2022pinpointing}.

To construct \(\mathcal{K}_{\text{ngram}}\), \methodname extracts unigrams, bigrams, and trigrams from EventId sequences in the training sessions.
Each candidate pattern is counted at the session level in anomalous and normal sessions.
Anomalous-session and normal-session support are calculated as the frequencies of anomalous and normal training sessions that contain the pattern, respectively.
For each candidate, \textit{lift} measures anomalous-session support relative to normal-session support, and candidates are ranked by lift in the descending order.
Overall, \emph{anomalous-session support}, \emph{normal-session support}, and \emph{minimum lift} obtained from the training data determine the initial candidate pool.
The validation data is further used to remove patterns that fire excessively on normal validation sessions or have low validation F1-score.
The remaining patterns form \(\mathcal{K}_{\text{ngram}}\).

During inference, \methodname first matches the EventId sequence \(E(x)\) of a new session against \(\mathcal{K}_{\text{ngram}}\).
A pattern fires if it appears as a contiguous subsequence of \(E(x)\).
If one or more patterns fire, the local event-pattern module produces an evidence signal \(s_n(x)=1\); otherwise, it produces \(s_n(x)=0\).
Formally,
\begin{align}
  s_n(x)=\mathbf{1}\{\exists p\in\mathcal{K}_{\text{ngram}}: p\sqsubseteq E(x)\}\label{eq:sn_def}
\end{align}
where \(p\sqsubseteq E(x)\) means that the EventId sequence pattern \(p\) appears as a contiguous subsequence of \(E(x)\).
The matched patterns are also recorded as explanation evidence together with their corresponding log templates.

\begin{table}
  \centering
  \caption{Example contiguous trigram ($n=3$) match in a log session.}
  \vspace{-0.5em}
  \label{tab:event_example}
  \footnotesize
  \renewcommand{\arraystretch}{1.12}
  \begin{tabular}{@{}c>{\raggedright\arraybackslash}p{0.72\columnwidth}@{}}
    \hline
    EventId & Log Template \\
    \hline
    $E_{22}$ & BLOCK* NameSystem.addStoredBlock: blockMap updated: \textless{}*\textgreater{} is added to \textless{}*\textgreater{} size \textless{}*\textgreater{} \\
    $E_{29}$ & PacketResponder \textless{}*\textgreater{} Exception java.io.EOFException \\
    $E_{41}$ & Received block \textless{}*\textgreater{} of size \textless{}*\textgreater{} from \textless{}*\textgreater{} \\
    $E_{41}$ & Received block \textless{}*\textgreater{} of size \textless{}*\textgreater{} from \textless{}*\textgreater{} \\
    $E_{44}$ & Deleting block \textless{}*\textgreater{} file \textless{}*\textgreater{} \\
    \hline
  \end{tabular}
  \vspace{-2em}
\end{table}

Table~\ref{tab:event_example} shows an example of deterministic local event-pattern matching.
The session is represented by the ordered event sequence \([E_{22}, E_{29}, E_{41}, E_{41}, E_{44}]\).
For each stored pattern \(p=(e_1,\ldots,e_n)\in\mathcal{K}_{\text{ngram}}\), the module slides a length-\(n\) window over the session and fires when any window is exactly equal to \(p\).
In this example, the length-3 window beginning at the second event is \((E_{29},E_{41},E_{41})\).
If this trigram is present in \(\mathcal{K}_{\text{ngram}}\), the module outputs \(s_n(x)=1\) and reports the fired trigram as the supporting evidence.

\subsubsection{\textbf{Cluster-Derived Rule Bank Construction}}

\label{sec:cluster_rule_detector}

\begin{figure}[t]
  \centering
  \includegraphics[width=0.7\columnwidth]{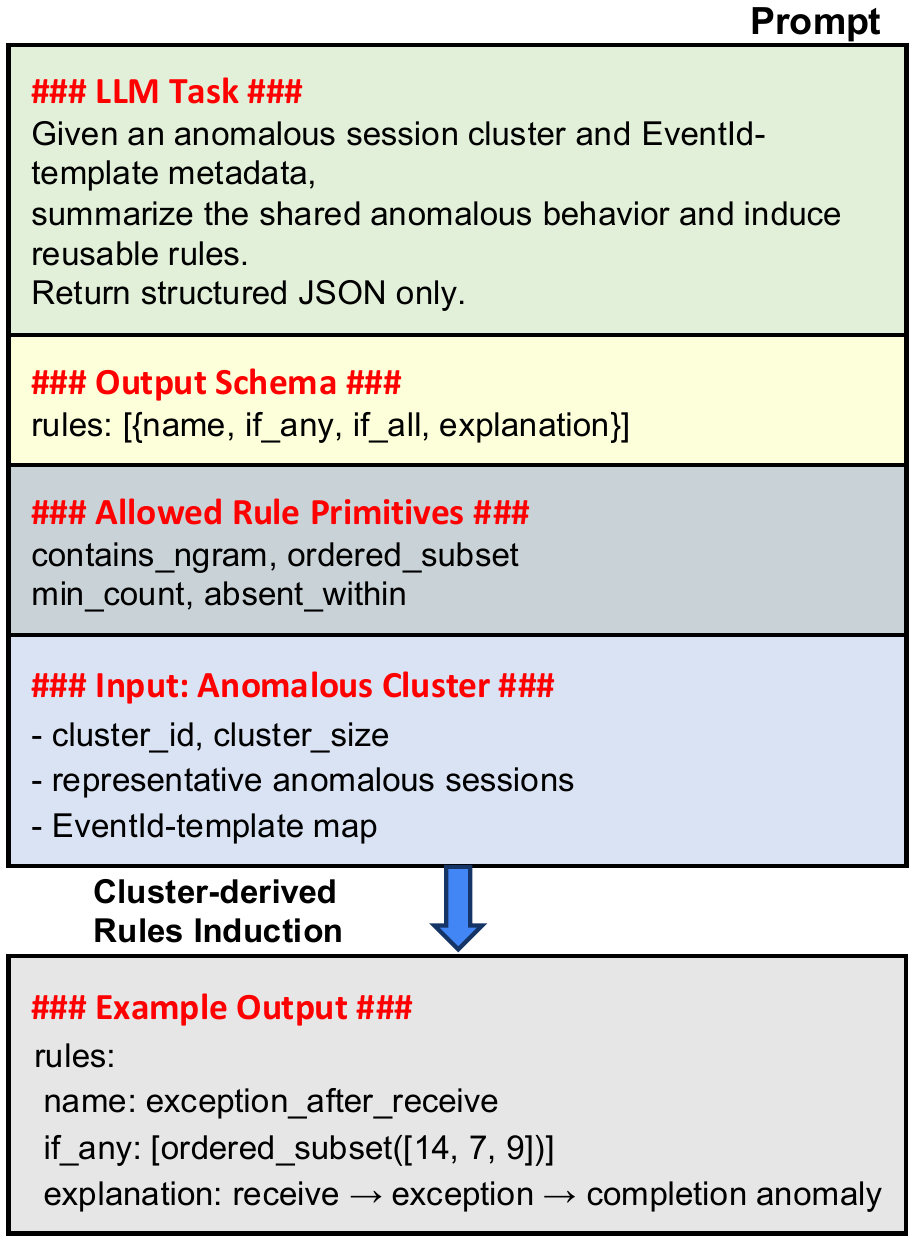}

  \caption{Example cluster-level prompt and LLM response for cluster-derived rule generation.}
  \label{fig:rule}
  \vspace{-1em}
\end{figure}

Local event patterns capture short contiguous behaviors, but many log anomalies involve broader operational patterns.
Examples include related events that occur non-contiguously, repeated state transitions, missing expected steps, or abnormal combinations of event counts.
To capture these behaviors, \methodname constructs the cluster-derived rule bank \(\mathcal{K}_{\text{rule}}\) from anomalous training sessions.

We first representeach anomalous training session as a Term Frequency--Inverse Document Frequency (TF-IDF) vector
over its EventId sequence~\cite{salton1988termweighting}.
We then apply HDBSCAN~\cite{mcinnes2017accelerated} to group sessions with similar event-distribution patterns.
Clustering-based log analysis has been shown to be effective and efficient for log-related tasks such as probabilistic label estimation, problem identification, and log-key abstraction~\cite{yang2021plelog,lin2016logcluster,egersdoerfer2023clusterlog}.
We chose HDBSCAN for three reasons.
First, the number of recurring anomaly behaviors is not known in advance, so methods that require a predefined number of clusters, such as K-means~\cite{hartigan1979algorithm}, are less suitable.
Second, HDBSCAN is density-based and can identify stable groups while treating sparse or idiosyncratic sessions as noise.
Third, HDBSCAN has a small number of user-facing parameters and has been reported to be robust to parameter selection in practice~\cite{mcinnes2017hdbscan}.
This allows \methodname to summarize recurring anomaly behaviors at the cluster level rather than generating rules from isolated sessions.

For each cluster, \methodname selects representative sessions and provides them to an LLM together with their EventId sequences, the EventId-template mapping, and basic cluster statistics.
Figure~\ref{fig:rule} shows the cluster-level prompt and an example structured response used for rule generation.
The LLM is prompted to produce two outputs: a human-readable summary of the recurring operational behavior and structured rule candidates that describe conditions associated with this behavior.
The generated rules are normalized into a constrained rule format over EventId sequences.
Each rule contains a name, an explanation, and one or more conditions built from a fixed set of primitives: \texttt{contains\_ngram}, \texttt{ordered\_subset}, \texttt{min\_count}, and \texttt{absent\_within}.
These primitives check whether a contiguous event subsequence appears, whether events appear in a specified order, whether an event appears at least a specified number of times, and whether an expected event is absent within a local window.

A rule is represented as structured output:
\texttt{\{name, if\_any, if\_all, explanation\}}.
A rule is considered fired on session \(x\) when all clauses in \texttt{if\_all} are satisfied and, if \texttt{if\_any} is non-empty, at least one clause in \texttt{if\_any} is satisfied.
If one or more rules fire, the cluster-derived rule bank matcher produces an evidence signal \(s_r(x)=1\); otherwise, it produces \(s_r(x)=0\).
Formally,
\begin{align}
  s_r(x)=\mathbf{1}\{\exists r\in\mathcal{K}_{\text{rule}}: r(E(x))=1\}\label{eq:sr_def}
\end{align}
The triggered rules and their cluster-level explanations are retained as evidence for explaining why the session was flagged.

Before a rule is stored in \(\mathcal{K}_{\text{rule}}\), \methodname validates it against the rule schema, checks allowed primitives and type correctness, and ensures consistency with the cluster's template dictionary.
Rules that fail these validations are discarded.
At inference, each retained rule is evaluated directly in code against \(E(x)\), without invoking LLMs.
This design confines LLM involvement to the training-time rule discovery, making inference-time rule matching deterministic and efficient.

\subsection{Inference-Time LLM-Based Semantic Reasoning and Explanation}

For each \textit{\textbf{unique}} log session, \methodname uses an LLM to 1) generate a semantic explanation and 2) assess whether that behavior indicates a high-risk anomaly during inference.
The explanation complements the rules derived during training, while the anomaly assessment provides an additional signal for the final decision.

The LLM-based semantic reasoning module uses a fixed prompt containing the ordered event-template sequence of the target session. 
The prompt asks the LLM to identify the expected behavior, summarize the observed behavior, and determine whether the difference indicates a high-risk anomaly.
To make these outputs directly usable by \methodname, the prompt specifies a structured response containing \texttt{expected}, \texttt{observed}, \texttt{is\_high\_risk}, and \texttt{confidence}.
The \texttt{expected} and \texttt{observed} fields form the semantic explanation of the session, whereas \texttt{is\_high\_risk} and \texttt{confidence} provide the assessment used in fusion.
The module is invoked once for each incoming session and does not require a separate training stage.

Given a new session \(x\), the module returns \texttt{is\_high\_risk}, confidence \(\gamma(x)\), and auxiliary rationale descriptors. The fusion signal from this module is defined as \begin{align} s_\ell(x)=\mathbf{1}\{\texttt{is\_high\_risk}=1 \land \gamma(x)\ge\tau_\ell\}\label{eq:sllm_def} \end{align} where \(\tau_\ell\) is selected on the validation split. If \(\gamma(x)<\tau_\ell\), the LLM-based semantic reasoning signal is treated as inactive for fusion. The remaining structured outputs are retained as structured rationale support and are not directly used in decision fusion.

Note that \methodname calls an LLM for every new log session and stores the result in a knowledge bank. For a session that has occurred in the past, it reuses the generated explanation from the knowledge bank without any LLM inferences.

\subsection{Evidence-Guided Inference}
\label{sec:inference}

The final knowledge bank \(\mathcal{K}\) contains \(\mathcal{K}_{\text{ngram}}\) and \(\mathcal{K}_{\text{rule}}\) obtained from the training stage, as well as the cache of session-level LLM-based semantic reasoning results.
Note that \(\mathcal{K}\) and backbone detectors remain static at test time and will not be updated.

\subsubsection{Evidence Matching and Reuse}
\label{sec:mathing}
An incoming session \(x\) is represented by its session identifier, ordered EventId sequence \(E(x)\), and aligned log templates.
The unchanged backbone produces an anomaly score \(a_b(x)\), which is converted into the binary backbone signal
\begin{align}
  s_b(x)=\mathbf{1}\{a_b(x)\ge\tau_b\}\label{eq:sb_def}
\end{align}
where \(\tau_b\) is threshold selected on the validation split.

The local event-pattern matcher checks contiguous subsequences of \(E(x)\) against \(\mathcal{K}_{\text{ngram}}\) to obtain \(s_n(x)\) and the matched patterns.
The cluster-derived rule bank matcher executes each retained rule in \(\mathcal{K}_{\text{rule}}\) against \(E(x)\) to obtain \(s_r(x)\) and the triggered rules.

The session-result cache is indexed by the exact session identifier.
On a cache hit, \methodname reuses the stored LLM-based semantic reasoning label, confidence, and rationale to obtain \(s_\ell(x)\).
On a cache miss, \methodname invokes the LLM-based semantic reasoning module and stores its structured result without using a test label.
The cache supports reuse but does not modify existing artifact bank.

\subsubsection{Inference-Time Four-Signal Fusion}
\label{sec:fusion}

For every session, \methodname obtains all four binary signals without priority ordering or early stopping.
The final session-level anomaly decision is
\begin{align}
  \hat{y}(x)=\mathcal{F}\big(s_b(x), s_n(x), s_r(x), s_\ell(x)\big)\label{eq:fusion_def}
\end{align}
where \(\mathcal{F}\) is selected from OR, AND, and vote-\(k\) policies.
The fusion policy, vote-\(k\), \(\tau_b\), and \(\tau_\ell\) are selected based on the validation data.

\subsubsection{Evidence-Based Explanation Generation}
\label{sec:evidence_based_outputs}

After fusion, \methodname generates an explanation by tracing which evidence sources contributed to the final anomaly decision.
It first collects the matched local event patterns and maps their EventIds back to the corresponding log templates so the explanation can refer to human-readable log behavior.
It then collects the triggered cluster-derived rules and attaches the cluster-level explanations from which those rules were compiled.
Finally, it adds the LLM-based semantic reasoning result, including its label, confidence, and rationale, as session-specific semantic context.
The resulting explanation combines the final label, backbone score or label, matched local event patterns, triggered rule explanations, and LLM-based rationale into one evidence trace.
Because cluster-level explanations are compiled into executable rules, they can affect future decisions and are returned whenever the corresponding rules fire.

\section{Experimental Setup}
\label{sec:setup}

\subsection{Datasets}
\label{sec:datasets}

\begin{table}
  \centering
  \small
  \begin{threeparttable}
  \caption{Overall statistics of evaluated datasets.}
  \label{tab:datasets}
\begin{tabular}{lrrr}
\toprule
\textbf{System} &
\textbf{\# Logs} &
\textbf{\# Sessions} &
\textbf{\# Anom. Sessions} \\
\midrule
HDFS    & 11,175,629 & 166,241 & 10,381 (6.24\%) \\
BGL     & 4,188,777  & 41,385  & 27,224 (65.78\%) \\
Liberty & 550,000    & 1,589   & 260 (16.35\%) \\
\bottomrule
\end{tabular}
\begin{tablenotes}
\item \footnotesize{\textbf{Note:} Values in parentheses indicate anomaly percentages.}
\end{tablenotes}
\end{threeparttable}
\vspace{-1.5em}
\end{table}

We evaluate \methodname on three public Loghub benchmarks: HDFS, BGL, and Liberty~\cite{zhu2023loghub}.
These benchmarks cover distributed file-system and supercomputer logs, including more than 20 million logs (Table~\ref{tab:datasets}).
We use the public structured releases of the entire HDFS and BGL.
For Liberty, we use a contiguous time-ordered slice of 550,000 log lines from the original stream.
This follows the common practice in log anomaly detection studies of evaluating public datasets under fixed, reproducible preprocessing protocols while keeping large-scale experiments tractable~\cite{le2021neuralog,guan2024logllm}.
Across all datasets, log entries are sorted by timestamp before session construction.

\noindent\textbf{Log session construction and split.}
Similar to prior studies~\cite{du2017deeplog,meng2019loganomaly}, log sessions are the unit of detection and evaluation.
HDFS and BGL events are grouped by block identifier and compute node, respectively. HDFS uses the official block-level labels, whereas a BGL session is considered anomalous if any session event is anomalous.
Liberty events are grouped into host-based temporal buckets, with a bucket labeled anomalous if it contains any anomalous event.

Following prior log anomaly detection evaluations, sessions are partitioned into training, validation, and test subsets using fixed 70\%/10\%/20\% ratios~\cite{le2021neuralog,guan2024logllm}.
The split is performed at the session level to prevent logs from the same session from contributing data to multiple partitions.
After sorting sessions by timestamp, we use the earliest 70\% for training, the next 10\% for validation, and the latest 20\% for testing, so no future sessions are used to train or tune models evaluated on earlier sessions.
Final metrics are reported at the session level after aggregating window-level predictions.

\noindent\textbf{Leakage prevention.}
During training, the backbone is trained on the training set using its original pipeline, while \methodname mines local event patterns and compiles cluster-level explanations of anomalous-session clusters into executable rules.
At test time, the resulting patterns and rules remain read-only in the \(\mathcal{K}\).
For each test session, the LLM-based semantic reasoning module retrieves a cached result when available or generates an anomaly prediction, confidence score, and rationale without access to the ground-truth label.

\subsection{Evaluation Metrics}
\label{sec:metrics}

We evaluate anomaly detection effectiveness using Precision, Recall, and F1-score at the session level.
Precision measures the fraction of predicted anomalies that are truly anomalous.
Recall measures the fraction of true anomalies that are successfully detected.
F1-score is the harmonic mean of Precision and Recall.

\subsection{Experimental Settings}
\label{sec:impl}

We implement \methodname with Python 3.12 using a unified pipeline for log parsing, session construction, backbone training, local event-pattern mining, anomaly clustering, LLM-based semantic reasoning, cluster-derived rule bank, and augmented inference.
We conduct experiments on a MacBook Pro (Apple M5, 10-core CPU, 16 GB RAM) running macOS 26.5.1.
Unless otherwise stated, window-based experiments use window size $w=20$ and stride $s=1$.
The evaluated backbone detectors are DeepLog~\cite{du2017deeplog}, LogAnomaly~\cite{meng2019loganomaly}, LogBERT~\cite{guo2021logbert}, and NeuralLog~\cite{le2021neuralog}.
We use the default model settings for each backbone.
We keep preprocessing, split construction, and evaluation metrics fixed across the baseline and \methodname variants.

For semantic reasoning, \methodname uses gpt-4o-mini-2024-07-18 through an API-based prompting workflow~\cite{openai2024gpt4omini}.
The LLM is used for session-level semantic reasoning and for cluster-level explanation generation.
Cluster-level explanations are compiled into explanation-derived executable rules and stored in the cluster-derived rule bank \(\mathcal{K}_{\text{rule}}\) for later reuse.
LLM outputs are requested in structured JSON format so labels, confidence scores, rationales, and rule candidates can be parsed consistently.
All LLM calls use a temperature of 0.

\section{Evaluation}
\label{sec:evaluation}

\subsection*{RQ1: How effective is \methodname for anomaly detection?}
\label{sec:rq1}

\noindent\textbf{Motivation.}
The primary goal of \methodname is to improve anomaly detection by augmenting an existing backbone detector with semantic evidence rather than replacing the backbone.
This RQ evaluates whether local event patterns, LLM-based semantic reasoning signals, and explanation-derived executable rules improve the final session-level decision across different datasets and backbone architectures.

\noindent\textbf{Approach.}
We compare each standalone backbone detector with its \methodname-augmented variant.
The baseline and \methodname-augmented variants use the same dataset split, preprocessing procedure, windowing configuration, and session-level evaluation metrics.
The backbone score remains the starting point, while \methodname adds test-time fusion over local event-pattern evidence, LLM-based semantic reasoning evidence, and explanation-derived executable rule evidence.

\noindent\textbf{Results.}
Table~\ref{tab:logxplain_comparison} reports Precision, Recall, F1-score, and F1 improvement for all dataset--backbone combinations using \methodname.

\begin{table}[t]
  \centering
  \caption{Baseline vs.\ \methodname\ augmented performance comparison.}
  \label{tab:logxplain_comparison}
  \setlength{\tabcolsep}{5pt}
  \renewcommand{\arraystretch}{1.15}
  \resizebox{\columnwidth}{!}{
   \begin{tabular}{ll|ccc|ccc|c}
    \toprule
    \textbf{Dataset}
    & \textbf{Backbone}
    & \multicolumn{3}{c|}{\textbf{Baseline}}
    & \multicolumn{3}{c|}{\textbf{\methodname}}
    & \textbf{$\Delta$F1} \\
    \cmidrule(lr){3-5}\cmidrule(lr){6-8}
    & & \textbf{P} & \textbf{R} & \textbf{F1} & \textbf{P} & \textbf{R} & \textbf{F1} & \\
    \midrule
    \multirow{4}{*}{HDFS}
    & LogBERT    & 0.92 & 0.79 & 0.85 & 1.00 & 0.98 & \textbf{0.99} & +0.14 \\
    & NeuralLog  & 0.83 & 0.87 & 0.85 & 0.83 & 1.00 & \textbf{0.91} & +0.06 \\
    & DeepLog    & 0.84 & 0.87 & 0.85 & 0.84 & 1.00 & \textbf{0.91} & +0.06 \\
    & LogAnomaly & 0.85 & 0.88 & 0.87 & 0.85 & 1.00 & \textbf{0.92} & +0.05 \\
    \midrule
    \multirow{4}{*}{BGL}
    & LogBERT    & 0.66 & 1.00 & 0.79 & 1.00 & 1.00 & \textbf{1.00} & +0.21 \\
    & NeuralLog  & 1.00 & 0.91 & 0.95 & 1.00 & 1.00 & \textbf{1.00} & +0.05 \\
    & DeepLog    & 1.00 & 0.93 & 0.96 & 1.00 & 1.00 & \textbf{1.00} & +0.04 \\
    & LogAnomaly & 1.00 & 0.95 & 0.98 & 1.00 & 1.00 & \textbf{1.00} & +0.03 \\
    \midrule
    \multirow{4}{*}{Liberty}
    & LogBERT    & 0.29 & 0.89 & 0.44 & 1.00 & 0.98 & \textbf{0.99} & +0.55 \\
    & NeuralLog  & 0.69 & 0.65 & 0.67 & 0.77 & 0.98 & \textbf{0.86} & +0.19 \\
    & DeepLog    & 0.91 & 0.60 & 0.72 & 0.94 & 0.98 & \textbf{0.96} & +0.24 \\
    & LogAnomaly & 1.00 & 1.00 & 1.00 & 1.00 & 1.00 & \textbf{1.00} & +0.00 \\
    \bottomrule
\end{tabular}
}
\vspace{-1.5em}
\end{table}

\noindent\textbf{\em \methodname improves F1-score whenever the baseline leaves room for improvement.}
Across the evaluated dataset--backbone combinations, every non-perfect baseline gains F1-score after augmentation.
The only unchanged case is LogAnomaly on Liberty, where the standalone backbone already achieves perfect F1-score.
The largest improvement occurs for LogBERT on Liberty, where F1-score increases from 0.44 to 0.99.
Large gains also appear for DeepLog on Liberty (+0.24), LogBERT on BGL (+0.21), and NeuralLog on Liberty (+0.19).
These results indicate that semantic augmentation is most useful when the backbone alone misses behavior that is visible in higher-level semantic evidence.

\noindent\textbf{\em The primary effectiveness gain comes from recovering previously missed anomalies.}
Recall reaches 1.00 for all BGL backbones and for NeuralLog, DeepLog, and LogAnomaly on HDFS.
LogBERT on HDFS also exhibits a substantial improvement, with recall increasing from 0.79 to 0.98.
On Liberty, recall improves to 0.98 for LogBERT, NeuralLog, and DeepLog.
These results indicate that \methodname effectively reduces false negatives while largely preserving the high precision of the underlying detectors.

A representative example can be observed in the Liberty dataset, where an anomalous session contains only five PBS scheduler failure messages (\texttt{pbs\_mom: task\_check, cannot tm\_reply to \ldots}) among 49 total log events, while the remaining messages correspond to benign SSH and session-management activity.
Because these failures are sparse, the backbone assigns a mean session score of 0.48, below the validation-tuned threshold of 0.55, and classifies the session as normal.
In contrast, \methodname correctly identifies the anomaly by leveraging semantic evidence capturing the recurring PBS scheduler-failure template \texttt{pbs\_mom: task\_check, cannot tm\_reply to <job>.ladmin2 task 1}.
Although the failure events constitute only a small minority of the session, their repeated occurrence across multiple PBS jobs forms a distinctive semantic pattern that is captured by the rule bank despite being diluted by the predominantly benign activity.
This example illustrates how semantic reasoning complements sequence-based detection by elevating semantically meaningful but infrequent anomaly indicators.

Overall, \methodname recovers 2,056 of 2,081 backbone false-negative prediction cases (98.8\%).
This includes 42 of 45 cases on Liberty, 900 of 917 on HDFS, and 1,114 of 1,119 on BGL across the four evaluated backbones.

\rqboxc{\methodname improves every non-perfect baseline and preserves the already perfect case, showing that semantic evidence can strengthen existing model-based log anomaly detectors without sacrificing precision or recall.}

\subsection*{RQ2: How efficient is \methodname in runtime and LLM usage?}
\label{sec:rq2}

\noindent\textbf{Motivation.}
LLM-assisted log analysis can become expensive if every session requires a fresh model call.
For a practical semantic augmentation framework, semantic reasoning should be acquired once, stored in the knowledge bank \(\mathcal{K}\), and reused through local event-pattern matching and executable rule evaluation whenever similar patterns recur.

\noindent\textbf{Approach.}
We evaluate efficiency from two perspectives.
First, we measure backbone training time, baseline inference runtime, and \methodname inference runtime on the same test partitions.
Second, we analyze LLM usage in terms of token consumption rather than monetary spending.
\methodname reduces repeated LLM dependence by storing local event patterns and explanation-derived executable rules in the knowledge bank.
We measure token usage separately for session-level prompting and cluster-level rule generation because these two stages have different context lengths and reuse patterns.

\noindent\textbf{Results.}
Table~\ref{tab:runtime_latency} reports the time cost of backbone training, \methodname training with one-time semantic-artifact construction, baseline inference, and \methodname inference on different backbones and datasets.

\begin{table}[t]
\centering
\caption{Comparison of training and inference time between the standalone baseline and \methodname.}
\vspace{-0.5em}
\label{tab:runtime_latency}
\footnotesize
\renewcommand{\arraystretch}{1.12}
\setlength{\tabcolsep}{3pt}

\begin{threeparttable}
\resizebox{\columnwidth}{!}{
\begin{tabular}{llrrrr}
\toprule
\textbf{Dataset}
& \textbf{Backbone}
& \multicolumn{2}{c}{\textbf{Training (s)}}
& \multicolumn{2}{c}{\textbf{Inference (s)}} \\
\cmidrule(lr){3-4}
\cmidrule(lr){5-6}
&
& \multicolumn{1}{c}{\textbf{Baseline}} & \multicolumn{1}{c}{\textbf{\methodname}} & \multicolumn{1}{c}{\textbf{Baseline}} & \multicolumn{1}{c}{\textbf{\methodname}} \\
\midrule

\multirow{4}{*}{HDFS}
& DeepLog    & 21.27  & 2415.93 & 5.16   & 10.97 {\scriptsize $(+5.81)$} \\
& NeuralLog  & 187.36 & 2527.30 & 39.03  & 44.66 {\scriptsize $(+5.63)$} \\
& LogAnomaly & 76.82  & 2381.75 & 17.39  & 23.12 {\scriptsize $(+5.73)$} \\
& LogBERT    & 603.98 & 2935.40 & 132.14 & 137.94 {\scriptsize $(+5.80)$} \\
\midrule

\multirow{4}{*}{BGL}
& DeepLog    & 72.09   & 347.29  & 16.80  & 20.73 {\scriptsize $(+3.93)$} \\
& NeuralLog  & 614.74  & 722.10  & 126.66 & 130.73 {\scriptsize $(+4.07)$} \\
& LogAnomaly & 262.31  & 368.85  & 54.14  & 58.21 {\scriptsize $(+4.07)$} \\
& LogBERT    & 2026.28 & 2148.68 & 442.67 & 446.73 {\scriptsize $(+4.06)$} \\
\midrule

\multirow{4}{*}{Liberty}
& DeepLog    & 31.13  & 288.04 & 1.12  & 3.48 {\scriptsize $(+2.36)$} \\
& NeuralLog  & 98.24  & 346.28 & 7.91  & 10.30 {\scriptsize $(+2.39)$} \\
& LogAnomaly & 20.10  & 272.08 & 3.38  & 5.79 {\scriptsize $(+2.41)$} \\
& LogBERT    & 161.43 & 322.75 & 12.46 & 14.83 {\scriptsize $(+2.37)$} \\
\bottomrule
\end{tabular}
}
\begin{tablenotes}[flushleft]
\footnotesize
\item\textbf{Note:} Values in parentheses indicate the additional \methodname inference time added to the backbone inference time.
\end{tablenotes}
\end{threeparttable}
\vspace{-1.5em}
\end{table}

\noindent\textbf{\em \methodname adds one-time training-stage overhead over the backbone.}
Compared with the standalone backbone, \methodname increases training-stage time because it additionally mines local event patterns, clusters anomalous training sessions, generates cluster-level semantic explanations, and compiles them into executable rules.
This extra cost is incurred only once during training-stage knowledge construction.
After construction, the resulting semantic artifacts are reused during test-time inference rather than regenerated for every session.

\noindent\textbf{\em \methodname adds modest and stable inference-time overhead over the backbone.}
Compared with standalone baseline inference, \methodname increases inference time because each session is additionally checked against reusable semantic artifacts.
The added time is stable in absolute terms: 5.63--5.81 seconds on HDFS, 3.93--4.07 seconds on BGL, and 2.36--2.41 seconds on Liberty.
This modest and stable overhead is useful in practice because semantic augmentation adds a mostly fixed local-processing cost rather than a backbone-dependent or LLM-call-dominated cost.
The stability comes from local knowledge-bank lookup, local event-pattern matching, executable rule evaluation, and reuse of cached LLM-based semantic reasoning results when available.
Thus, \methodname pays a limited per-session inference overhead to obtain semantic evidence and explanations, instead of repeatedly reconstructing semantic knowledge during testing.

\noindent\textbf{\em LLM cost is amortized through knowledge reuse.}
Across the full evaluation, \methodname uses 286M, 17M, and 13M LLM tokens for HDFS, BGL, and Liberty, respectively.
Session-level requests use approximately 624 tokens on average, including model outputs, while cluster-level requests typically use 2,500--4,000 tokens.
However, \methodname invokes the LLM only for previously unseen session- or cluster-level patterns and stores the resulting explanations and rules in the knowledge bank for subsequent reuse.
Therefore, its LLM cost scales with the number of novel patterns rather than the total number of processed sessions, requiring substantially fewer calls than a pipeline that invokes an LLM for every session.

\rqboxc{Compared with standalone backbones, \methodname adds training-stage and inference-time cost. The training-stage cost is incurred once to construct reusable semantic artifacts, while the inference overhead remains modest and stable because most added operations are local. Through knowledge-bank reuse, LLM usage scales with newly observed session or cluster patterns rather than total session volume.}

\subsection*{RQ3: How useful are \methodname explanations?}
\label{sec:rq3}

\noindent\textbf{Motivation.}
An anomaly detector is more useful to developers when it explains why a session was flagged.
\methodname treats explanations as detection evidence rather than only post-hoc text, because cluster explanations are compiled into executable rules and stored in the cluster-derived rule bank \(\mathcal{K}_{\text{rule}}\).

\noindent\textbf{Approach.}
We evaluate explanation quality through a user study.
The study compares explanations generated by \methodname against explanations generated directly by a general-purpose LLM for the same anomalous sessions.
For the user study, one author randomly selected 10 anomalous test sessions and collected paired explanations from \methodname and GPT-4o-mini.
We recruited 14 participants through several software engineering research mailing lists.
The participant pool consisted of nine CS/SE PhD students, two postdoctoral researchers, and three professional software developers.
Nine participants self-reported more than five years of software development experience, while four reported between two and five years.
Participants also reported varying levels of log-analysis experience, with five reporting less than one year of experience, two reporting one year of experience, four reporting between two and five years, and two reporting more than five years of experience.

For each anomalous session, participants were presented with anonymized \methodname{} and GPT-4o-mini explanations.
The order of sessions and explanations was randomized and counterbalanced across participants.
They were asked to evaluate both explanations in terms of their \textit{logical coherence}, \textit{technical soundness}, \textit{label alignment}, \textit{clarity}, and \textit{overall quality} on a five-point Likert scale~\cite{likert1932technique} (1---Strongly Disagree, 5---Strongly Agree).

\noindent\textbf{Results.}
Table~\ref{tab:human_eval} reports the human evaluation results.

\begin{table}[t]
\centering
\caption{Human evaluation of explanation quality.}
\vspace{-0.5em}
\label{tab:human_eval}
\setlength{\tabcolsep}{7pt}
\renewcommand{\arraystretch}{1.1}
\begin{threeparttable}
\resizebox{0.7\columnwidth}{!}{
\begin{tabular}{lcc}
\toprule
\textbf{Metric} & \textbf{\methodname} & \textbf{Baseline LLM} \\
\midrule
Logical Coherence
& \textbf{4.19 $\pm$ 0.93}
& 3.35 $\pm$ 0.87 \\

Technical Soundness
& \textbf{4.06 $\pm$ 1.03}
& 3.17 $\pm$ 0.88 \\

Label Alignment
& \textbf{4.15 $\pm$ 0.98}
& 3.16 $\pm$ 0.82 \\

Clarity
& \textbf{4.35 $\pm$ 0.80}
& 3.07 $\pm$ 0.96 \\

Overall Quality
& \textbf{4.13 $\pm$ 0.98}
& 3.06 $\pm$ 0.84 \\
\bottomrule
\end{tabular}
}
\begin{tablenotes}
\item \footnotesize{\textbf{Note:} Values are reported as mean $\pm$ standard deviation. A higher value indicates better user perception.}
\end{tablenotes}
\end{threeparttable}
\vspace{-1.5em}
\end{table}

\noindent\textbf{\em Human evaluators consistently prefer \methodname explanations over direct LLM explanations.}
\methodname receives higher mean scores on all five dimensions.
The largest difference appears in Clarity, where \methodname scores 4.35 compared with 3.07 for the baseline LLM.
Logical Coherence, Technical Soundness, Label Alignment, and Overall Quality also improve by about one point or more on the five-point scale.
A paired Wilcoxon signed-rank test~\cite{wilcoxon1945individual} shows that all five differences are statistically significant ($p<0.001$). These quantitative improvements are also reflected in participants' qualitative feedback. For example, one participant noted that ``\textit{Reasoning A (\methodname) is more informative and better grounded [than the baseline].}'' Another participant also commented that ``\textit{Reasoning B (\methodname) is stronger than Reasoning A (baseline) because it is more specific to HDFS behavior and better connects the empty packet, invalid block cleanup, and interruption of block transfer.}''

\noindent\textbf{\em Cluster-level context is one plausible explanation for the higher human ratings.}
Figure~\ref{fig:hdfs_explanation_example} presents an HDFS explanation pair used in the human study.
The baseline LLM identifies a broad block-management issue but does not connect the events to a concrete failure mechanism.
\methodname identifies invalid block states, cleanup behavior, and replication pipeline failure as a coherent anomaly pattern.
This example is consistent with the possibility that aggregating related anomalous training sessions provides context for a more specific explanation, while the associated executable rule maintains an explicit connection to inference-time evidence.

\rqboxc{\methodname produces explanations that participants rate as clearer, more consistent, and more aligned with anomaly behavior than direct LLM explanations.}

\begin{figure}[t]
\centering
\fbox{
\begin{minipage}{0.94\columnwidth}
\footnotesize
\noindent\textbf{Raw log session.}\\
\texttt{BLOCK* NameSystem.delete: <*> is added to invalidSet <*>}\\
\texttt{Deleting block <*> file <*>}\\
\texttt{Redundant addStoredBlock request received <*>}

\vspace{3pt}
\noindent\textbf{Baseline LLM reasoning.}\\
The presence of redundant requests in a deletion sequence may indicate potential issues with block management or consistency within HDFS.

\vspace{3pt}
\noindent\textbf{\methodname reasoning.}\\
The session indicates a failure in the HDFS block replication pipeline.
Block metadata operations indicate inconsistent or invalid block state.
Blocks are added to the invalid set and delete or cleanup operations are triggered during replication.
As a result, the NameNode marks several blocks as invalid and triggers cleanup actions.
The combination of invalid block metadata and cleanup suggests a replication pipeline breakdown rather than normal block lifecycle operations.
\end{minipage}}
\vspace{-0.2em}
\caption{One explanation example from HDFS.}
\label{fig:hdfs_explanation_example}
\vspace{-1.5em}
\end{figure}

\subsection*{RQ4: How do different settings and training-data percentages affect \methodname performance?}
\label{sec:rq4}

\noindent\textbf{Motivation.}
\methodname combines backbone predictions, local event patterns, LLM-based semantic reasoning signals, and explanation-derived executable rules.
Understanding how these components contribute is necessary for interpreting the augmented decision process.
It is also important to test whether \methodname remains effective when less training data is available for backbone learning, local event-pattern mining, clustering, and knowledge bank construction.

\noindent\textbf{Approach.}
For the ablation study, we compare the complete \methodname framework with its variants by removing LLM-based semantic reasoning, local event patterns, or the cluster-derived rule bank while retaining the remaining signals.
For the sensitivity study, we vary only the amount of training data while keeping the same validation and test sets across all configurations.
We evaluate training proportions of 30\%, 50\%, and 70\% and report F1-score for each dataset--backbone combination using the fixed validation and test partitions.

\noindent\textbf{Results.}
Table~\ref{tab:ablation} reports the component ablation results, and Table~\ref{tab:split_sensitivity} compares each backbone with its \methodname-augmented variant at each training-set percentage.

\begin{table}[b]
  \vspace{-1em}
\centering

\caption{\methodname\ performance (F1 score) under different settings.}
\label{tab:ablation}
\vspace{-0.5em}
\setlength{\tabcolsep}{2.5pt}
\renewcommand{\arraystretch}{1.2}
\resizebox{\columnwidth}{!}{
\begin{tabular}{ll|c|ccc}
\toprule
\textbf{} &
\textbf{Backbone} &
\textbf{\methodname} &
\textbf{w/o LLM} &
\textbf{w/o Pattern} &
\textbf{w/o Rule} \\
\midrule
\multirow{4}{*}{HDFS}
& DeepLog
& 0.91
& 0.91 ($\downarrow$0.33\%)
& 0.86 ($\downarrow$6.13\%)
& 0.91 ($\downarrow$0.33\%) \\

& NeuralLog
& 0.91
& 0.91 (0.00\%)
& 0.86 ($\downarrow$5.82\%)
& 0.91 (0.00\%) \\

& LogAnomaly
& 0.92
& 0.91 ($\downarrow$1.09\%)
& 0.86 ($\downarrow$6.75\%)
& 0.91 ($\downarrow$0.76\%) \\

& LogBERT
& 0.99
& 0.97 ($\downarrow$2.53\%)
& 0.36 ($\downarrow$63.33\%)
& 0.81 ($\downarrow$18.38\%) \\

\midrule

\multirow{4}{*}{BGL}
& DeepLog
& 1.00
& 1.00 (0.00\%)
& 1.00 (0.00\%)
& 0.97 ($\downarrow$3.00\%) \\

& NeuralLog
& 1.00
& 1.00 (0.00\%)
& 1.00 (0.00\%)
& 0.95 ($\downarrow$5.20\%) \\

& LogAnomaly
& 1.00
& 1.00 (0.00\%)
& 1.00 (0.00\%)
& 0.97 ($\downarrow$2.70\%) \\

& LogBERT
& 1.00
& 1.00 (0.00\%)
& 0.56 ($\downarrow$43.90\%)
& 0.97 ($\downarrow$2.60\%) \\

\midrule

\multirow{4}{*}{Liberty}
& DeepLog
& 0.96
& 0.94 ($\downarrow$1.87\%)
& 0.71 ($\downarrow$26.72\%)
& 0.94 ($\downarrow$1.87\%) \\

& NeuralLog
& 0.86
& 0.84 ($\downarrow$2.89\%)
& 0.84 ($\downarrow$2.89\%)
& 0.84 ($\downarrow$2.89\%) \\

& LogAnomaly
& 1.00
& 1.00 (0.00\%)
& 1.00 (0.00\%)
& 1.00 (0.00\%) \\

& LogBERT
& 0.99
& 0.90 ($\downarrow$9.09\%)
& 0.83 ($\downarrow$16.16\%)
& 0.83 ($\downarrow$16.16\%) \\

\bottomrule
\end{tabular}}
\par\vspace{1mm}
\noindent
\begin{minipage}{\columnwidth}
\footnotesize\textbf{Note:} LLM, Pattern, and Rule denote LLM-based semantic reasoning, local event patterns, and the cluster-derived rule bank, respectively.
Values in parentheses show percentage differences from the full \methodname.
\end{minipage}
\vspace{-1.5em}
\end{table}

\begin{table}[t]
\centering
\caption{Baseline and \methodname-augmented F1-score under different training-set percentages.}
\vspace{-0.5em}

\label{tab:split_sensitivity}
\small
\setlength{\tabcolsep}{3.5pt}
\renewcommand{\arraystretch}{1.1}
\resizebox{\columnwidth}{!}{
\begin{tabular}{llcccccc}
\toprule
\textbf{} & \textbf{Backbone} & \multicolumn{2}{c}{\textbf{30\%}} & \multicolumn{2}{c}{\textbf{50\%}} & \multicolumn{2}{c}{\textbf{70\%}} \\
\cmidrule(lr){3-4}\cmidrule(lr){5-6}\cmidrule(lr){7-8}
& & \textbf{Base} & \textbf{\methodname} & \textbf{Base} & \textbf{\methodname} & \textbf{Base} & \textbf{\methodname}  \\
\midrule
\multirow{4}{*}{HDFS}
& DeepLog    & 0.88 & 0.93 ($\uparrow$5.68\%) & 0.87 & 0.93 ($\uparrow$6.90\%) & 0.85 & 0.91 ($\uparrow$7.06\%) \\
& NeuralLog  & 0.86 & 0.92 ($\uparrow$6.98\%) & 0.86 & 0.91 ($\uparrow$5.81\%) & 0.85 & 0.91 ($\uparrow$7.06\%) \\
& LogBERT    & 0.85 & 0.99 ($\uparrow$16.47\%) & 0.85 & 0.99 ($\uparrow$16.47\%) & 0.85 & 0.99 ($\uparrow$16.47\%) \\
& LogAnomaly & 0.87 & 0.93 ($\uparrow$6.90\%) & 0.88 & 0.93 ($\uparrow$5.68\%) & 0.87 & 0.92 ($\uparrow$5.75\%) \\
\midrule
\multirow{4}{*}{BGL}
& DeepLog    & 0.95 & 1.00 ($\uparrow$5.26\%) & 0.98 & 1.00 ($\uparrow$2.04\%) & 0.96 & 1.00 ($\uparrow$4.17\%) \\
& NeuralLog  & 0.95 & 1.00 ($\uparrow$5.26\%) & 0.98 & 1.00 ($\uparrow$2.04\%) & 0.95 & 1.00 ($\uparrow$5.26\%) \\
& LogBERT    & 0.79 & 1.00 ($\uparrow$26.58\%) & 0.79 & 1.00 ($\uparrow$26.58\%) & 0.79 & 1.00 ($\uparrow$26.58\%) \\
& LogAnomaly & 0.97 & 1.00 ($\uparrow$3.09\%) & 0.97 & 1.00 ($\uparrow$3.09\%) & 0.98 & 1.00 ($\uparrow$2.04\%) \\
\midrule
\multirow{4}{*}{Liberty}
& DeepLog    & 0.84 & 0.99 ($\uparrow$17.86\%) & 0.67 & 0.98 ($\uparrow$46.27\%) & 0.72 & 0.96 ($\uparrow$33.33\%) \\
& NeuralLog  & 0.76 & 0.90 ($\uparrow$18.42\%) & 0.62 & 0.89 ($\uparrow$43.55\%) & 0.67 & 0.86 ($\uparrow$28.36\%) \\
& LogBERT    & 0.60 & 0.99 ($\uparrow$65.00\%) & 0.41 & 0.95 ($\uparrow$131.71\%) & 0.44 & 0.99 ($\uparrow$125.00\%) \\
& LogAnomaly & 0.44 & 0.99 ($\uparrow$125.00\%) & 0.97 & 0.97 (0.00\%) & 1.00 & 1.00 (0.00\%) \\
\bottomrule
\end{tabular}}

\par\vspace{1mm}
\noindent
\begin{minipage}{\columnwidth}
\footnotesize\textbf{Note:} Values in parentheses indicate the percentage change relative to the baseline trained with the same training-set percentage.
\end{minipage}
\vspace{-1.5em}
\end{table}

\noindent\textbf{\em Local event patterns provide the strongest direct detection signal.}
Removing the local event-pattern module causes the largest degradation in most non-saturated settings.
The effect is particularly pronounced for LogBERT, whose F1-score decreases from 0.99 to 0.36 on HDFS, from 1.00 to 0.56 on BGL, and from 0.99 to 0.83 on Liberty.
This indicates that explicit local event combinations recover anomaly evidence not consistently captured by the backbone or the other semantic evidence modules.

\noindent\textbf{\em The cluster-derived rule bank provides reusable evidence for evidence-based explanations.}
Removing the rule bank lowers every BGL result from 1.00 to 0.95--0.97.
Its contribution is strongest for LogBERT, where F1-score decreases from 0.99 to 0.81 on HDFS and from 0.99 to 0.83 on Liberty.
For the other HDFS backbones, the change is at most 0.01, showing that the value of rule bank evidence depends on the backbone and dataset.

\noindent\textbf{\em LLM-based semantic reasoning provides smaller but targeted gains.}
Removing LLM-based semantic reasoning has no measurable effect on BGL and changes HDFS F1-score by at most 0.02.
Its largest effect occurs for LogBERT on Liberty, where F1-score decreases from 0.99 to 0.90.
No ablation improves over the complete framework, supporting the use of LLM-based semantic reasoning as complementary evidence rather than as the primary detector.

\noindent\textbf{\em Semantic augmentation remains effective with less training data.}
Across the 36 dataset--backbone--training-percentage combinations, \methodname improves the standalone backbone in 34 cases and ties it in the remaining two.
The relative gains range from 2.04\% to 26.58\% on BGL and from 5.68\% to 16.47\% on HDFS, while every augmented BGL configuration reaches an F1-score of 1.00.
The largest gains occur on Liberty, including improvements from 0.41 to 0.95 for LogBERT at 50\% training data and from 0.44 to 0.99 for LogAnomaly at 30\% training data.
The augmented results are also stable across training-set percentages: each HDFS backbone varies by at most 0.02, each Liberty backbone varies by at most 0.04, and all BGL backbones remain unchanged at 1.00.
These results show that the semantic augmentation consistently adds useful evidence over the backbone and retains its effectiveness when the training partition is reduced to 30\%.

\rqboxc{Local event patterns provide the largest direct performance contribution, the cluster-derived rule bank adds reusable evidence for evidence-based explanations, and LLM-based semantic reasoning supplies smaller but targeted gains in difficult configurations. Across training-set percentages, \methodname improves or matches every paired backbone baseline and remains stable with only 30\% training data.}

\section{Threats to Validity}
\label{sec:threats-to-validity}

\phead{Internal Validity.}
Our results may be affected by preprocessing, session construction, clustering, prompting, fusion choices, and the particular training split.
To reduce this threat, we use the same pipeline for each backbone and its \methodname-augmented variant, tune thresholds only on validation data, and perform local event-pattern mining, anomaly clustering, cluster-level explanation generation, and explanation-derived executable rule induction using only training data.
We further examine robustness under different training-set percentages, where \methodname remains effective when the available training data is reduced to 70\%, 50\%, and 30\%.
At test time, artifacts constructed from the training split remain read-only in the knowledge bank to prevent leakage.
Because the evaluated logs are publicly available, we cannot rule out that the pretrained LLM encountered benchmark-related content during pretraining, which may inflate its semantic reasoning performance.
The ablation results partly bound this threat because removing the LLM-based semantic reasoning signal still yields F1-scores from 0.84 to 1.00 across all evaluated settings.
This suggests that the main detection gains do not depend solely on possible public-benchmark memorization by the pretrained LLM.

\phead{External Validity.}
We evaluate \methodname on HDFS, BGL, and Liberty with multiple backbone detectors, but these benchmarks may not represent all production environments.
Real systems may contain evolving templates, unseen anomaly types, noisy labels, and cross-service failures.

\phead{Construct Validity.}
We use session-level Precision, Recall, and F1-score for detection and human ratings for explanation quality.
These metrics capture the main goals of \methodname, but do not fully measure alert fatigue, triage usefulness, within-session localization, or deployment cost.
The human study is also limited by its participants and sampled sessions.

\section{Conclusion}
\label{sec:conclusion}

In this paper, we presented \methodname, a plug-in semantic augmentation framework that preserves existing backbone detectors while adding semantic evidence from local event patterns, LLM-based semantic reasoning, and cluster-derived executable rules.
\methodname turns semantic interpretation into reusable evidence that can improve the final anomaly decision and explain why a session is flagged.
Across HDFS, BGL, and Liberty with four backbones, \methodname improves every non-perfect baseline, preserves the already perfect case, recovers most backbone false negatives, and produces explanations that users rate higher than direct LLM explanations.
These gains require only modest and stable inference-time overhead, suggesting that reusable semantic evidence can complement sequence-based detectors without making LLMs the sole detection mechanism.
Overall, the results show that semantic augmentation is a practical way to strengthen existing log anomaly detectors while making their decisions more interpretable and reusable.

\balance
\bibliographystyle{unsrt}
\bibliography{ref}
\end{document}